\documentstyle[12pt]{article}
\begin{document}
\title{Quantum Averaging I: Poincar\'e--von Zeipel
                            is Rayleigh--Schr\"odinger}
\author{Wolfgang Scherer \\ Institut f\"ur Theoretische Physik A \\
                           TU Clausthal \\ Leibnizstr. 10 \\
                           D--38678 Clausthal--Zellerfeld \\ Germany}
\date{}
\maketitle
\begin{abstract}
An exact analogue of the method of averaging in classical mechanics
is constructed for self--adjoint operators. It is shown to be
completely equivalent to the usual Rayleigh--Schr\"odinger
perturbation theory but gives
the sums over intermediate states in closed form expressions.
The anharmonic oscillator and the Henon--Heiles system
are treated as examples
to illustrate the quantum averaging method.
\end{abstract}
\section{Introduction and motivation}
\label{s1}
The failure to obtain exact solutions for most mechanical systems of
interest (e.~g.~planetary motion) has prompted the search for
perturbation techniques almost immediately after the conception of
Newtonian mechanics~(see~\cite{Ciacag} for some history on the subject).
At about the turn of the last century
Lindstedt~\cite{Lindst}, Poincar\'e~\cite{Poinca}, and later von
Zeipel~\cite{Zeipel} developed a perturbation method for classical
Hamiltonian systems using an averaging procedure in phase space.
Despite its lack of convergence in many cases this method which we
shall henceforth refer to as the Poincar\'e--von Zeipel method has been
a widely used one since it yields at least asymptotic expansions.

Concerning the ability to find exact solutions nothing much changed
with the advent of quantum mechanics. There it turned out to be equally
important to develop perturbation methods and this was done
simultaneously with the beginning of quantum mechanics by
Schr\"odinger~\cite{Schroe}. Due to previous contributions to similar
perturbation techniques in other wave equations by Lord Rayleigh
this theory has been named Rayleigh--Schr\"odinger perturbation theory
and has later been given a rigorous mathematical basis in the work
of Kato~\cite{Kato} and Rellich~\cite{Rellic}.

In this paper it is shown that the two methods are identical. More
precisely, it will be shown that in quantum mechanics an exact analogue
of the classical Poincar\'e--von Zeipel method can be formulated with
the help of an averaging techique for self adjoint operators analogous
to the classical method and that the resulting quantum
Poincar\'e--von Zeipel
perturbation theory is identical to the Rayleigh--Schr\"odinger theory.

The analogy between the cassical and quantum case is based entirely on
the structure of the equations appearing in the algorithm and the
strucure of the method used to solve them (averaging). The
starting point in the classical case is a Hamiltonian function
on phase space with a perturbing Hamiltonian, whereas in the
quantum case we start from a self adjoint Hamiltonian operator
with a perturbing operator. If and how the two Hamiltonians are
related is of no interest here. No quantization or other
quantum--classical map (e.~g.~semi--classical correspondence)
is needed to construct the quantum analogue of the classical
Poincar\'e--von Zeipel theory.

Viewing Rayleigh--Schr\"odinger as a quantum version of the classical
Poincar\'e--von Zeipel method yields (apart from a purely conceptual
viewpoint)
one possible advantage: it gives closed expressions for the
sums over intermediate states which appear in the corrective terms
for the eigenvalues and eigenfunctions.

The method of quantum averaging has also been used to construct
a quantum analogue of Kolmogorov's superconvergent
perturbation theory~\cite{Kolmog}. This new quantum ``superconvergent"
perturbation theory is substantially different and from the standard
Rayleigh--Schr\"odinger method and initial numerical studies in some
examples indicate
much better convergence properties~\cite{SUCO}. While using
quantum averaging to construct analogues of the classical
Poincar\'e--von Zeipel and superconvegent methods they needed to be
compared with
existing perturbation methods in quantum mechanics. In~\cite{SUCO}
we have shown that the quantum superconvergent method yields
a new kind of perturbation theory and in this paper we show
that the quantum Poincar\'e--von Zeipel method is identical to the
standard Rayleigh--Schr\"odinger theory.

In classical mechanics the Poincar\'e--von Zeipel series
is in most cases divergent and yields only an asymptotic series.
This is similar in its quantum equivalent the
Rayleigh--Schr\"odinger series.
Since we prove the equivalence of the two methods we shall not
state all technical details necessary to make all steps
rigorous but refer the reader to the vast mathematical literature
dealing with the Rayleigh--Schr\"odinger (or Kato--Rellich) perturbation
theory~(see e.~g.~\cite{ReeSim}).
Just as certain quantities diverge in standard
Rayleigh--Schr\"odinger theory
(in cases where convergence conditions fail) the
power expansions in $\epsilon$
formally written down here may not converge in which case
the perturbation algorithm gives only asymptotic information,
all sums have to be replaced by finite ones up to $N$, and
equations have to be read modulo $O(\epsilon^{N+1})$ for
any finite $N$.

The paper is organized as follows:

In section~\ref{s2} we present the classical
Poincar\'e--von Zeipel perturbation
theory and method of averaging in such a way that it can easily
be generalized to quantum mechnics which is done in section~\ref{s3}.

In section~\ref{s4} we apply the quantum
Poincar\'e--von Zeipel and averaging
method to a Hamiltonian with pure point spectrum, show that
up to second order all results from the Rayleigh--Schr\"odinger theory
are reproduced and discuss two examples which illustrate
the method and
show possible advantages of this new way of constructing
the Rayleigh--Schr\"odinger series.

In section~\ref{s5} the full equivalence of the two perturbation
expansions in all orders is proven.

Finally, in section~\ref{s6} we discuss previous constructions
mimicking classical perturbation expansions in quantum mechanics
by Kummer~\cite{Kumm71,Kumm93}, Ali~\cite{Ali}, Eckhardt~\cite{Eckhar},
and Ben Lemlih and Ellison~\cite{LemEll}
and their relation to the present method and conclude with some
remarks about future investigations.
\section{Classical Poincar\'e--von Zeipel perturbation theory}
\label{s2}
In this section we will describe the classical Poincar\'e--von Zeipel
perturbation theory along with the method of averaging in a geometric
manner such that its generalization to quantum mechanics is almost
self--evident. To avoid later confusion we will use lower case letters
for the classical situation. The unperturbed Hamiltonian $h_0$
is a function on phase space $\gamma$ which is equipped
with a Poisson bracket structure $\{\cdot ,\cdot \}$. $h_0$ is
assumed to be sufficiently smooth, integrable and nondegenerate
in the sense of Liouville--Arnold, i.~e.~it has
$n:=\frac{1}{2}dim\gamma$ functionally independent
constants of motion $b_1,\dots ,b_n$ which are in involution, define
the invariant tori and have the property that
\[
\{h_0,g\}=0\Rightarrow g=g(b_1,\dots ,b_n).
\]
 Furthermore, let
\begin{equation}
\label{e2.1}
h(\epsilon):=\sum_{p=0}^\infty \frac{\epsilon^p}{p!}h_p
\end{equation}
be the perturbed Hamiltonian where the perturbations $h_p,\; p\ge 1$
are assumed to be sufficiently smooth functions on $\gamma$.
The idea of Poincar\'e--von Zeipel perturbation theory is to look for an
$\epsilon$--dependent generating function
\begin{equation}
\label{e2.2}
w(\epsilon):=\sum_{p=0}^\infty \frac{\epsilon^p}{p!} w_{p+1}
\end{equation}
(with $\epsilon$--independent smooth functions $w_l$)
such that $-w(\epsilon)$ generates
a canonical flow $\xi(\epsilon)$ with ``time" $\epsilon$.
Then $\varphi(\epsilon):=\xi(\epsilon)^{-1}$ is a transformation on $\gamma$
determined uniquely by
\begin{eqnarray}
\frac{d}{d\epsilon} \varphi (\epsilon)^* & = &
ad\; w(\epsilon) \circ \varphi (\epsilon)^*   \label{awppa} \\
\varphi (0) & = & id_\gamma  \label{awppb}
\end{eqnarray}
and gives
rise to the following action on phase space functions $a$:
\begin{equation}
\label{e2.3}
\varphi(\epsilon)^*a:= a\circ \varphi(\epsilon) .
\end{equation}
Here
$ad\, f(g)$
is defined for any two phase space functions $f,g$  as
\begin{equation}
\label{e2.4}
ad\, f(g) := \{ f, g\}
\end{equation}
and for future use we remark that
$(ad\, w(\epsilon) )^p:= ad\, w(\epsilon)\circ\cdots\circ
ad\, w(\epsilon)$  ($p$ times).
For later purpose we shall need an expansion of $\varphi^*$ in terms
of some differential operators $t_p$ independent of $\epsilon$:
\begin{equation}
\label{e2.4a}
\varphi(\epsilon)^*=\sum_{p=0}^\infty\frac{\epsilon^p}{p!} t_p.
\end{equation}
The $t_p$ are then recursively defined through $t_0=id_\gamma$ and
\begin{equation}
\label{e2.4b}
t_{p+1}=\sum_{l=0}^p {p \choose l} ad\, w_{l+1} \circ t_{p-l}.
\end{equation}
Acting with $\varphi(\epsilon)^*$
on the perturbed Hamiltonian $h(\epsilon)$
gives a new Hamiltonian
\begin{equation}
\label{e2.5}
k(\epsilon):=\varphi(\epsilon)^* h(\epsilon)
\end{equation}
which is assumed to be analytic in $\epsilon$:
\begin{equation}
\label{e2.6}
k(\epsilon)=\sum_{p=0}^\infty\frac{\epsilon^p}{p!} k_p
\end{equation}
and for which one finds
\begin{eqnarray}
k_0 & = & h_0   \label{e2.6a} \\
k_p & = & ad\, w_p (h_0) + f_p,\qquad p\ge 1 \label{e2.6b}
\end{eqnarray}
where $f_1=h_1$ and for $p\ge 2$
\begin{equation}
\label{e2.6c}
f_p:= h_p +\sum_{l=0}^{p-2} {p-1 \choose l} (ad\, w_{l+1} (k_{p-l-1})
+t_{p-l-1} h_{l+1} ).
\end{equation}
If we had a solution $\psi_{k(\epsilon)}(t):\gamma\to\gamma$ of the
motion with Hamiltonian $k(\epsilon)$ then
\begin{equation}
\label{e2.7}
\psi_{h(\epsilon)}(t):= \varphi(\epsilon)\circ\psi_{k(\epsilon)}(t)
\circ\varphi(\epsilon)^{-1}
\end{equation}
would give us the desired solution for the perturbed Hamiltonian
$h(\epsilon)$. In general it is not possible to find $\varphi(\epsilon)$
such that a solution for $k(\epsilon)$ may be found. However, we may
choose the $w_l$ successively in such a manner that each $k_p$
is integrable, i.~e.~(since $h_0$ is integrable and nondegenerate)
such that
\begin{equation}
\label{e2.8}
ad\,h_0 (k_p)=0\qquad \forall p.
\end{equation}
In this way one can trivially solve the equations of motion for
$k(\epsilon)$ up to any finite order in $\epsilon$ and thus obtain
via~(\ref{e2.7}) solutions of $h(\epsilon)$ up to the same order
in $\epsilon$.

Thus we have to find the $w_p$ successively such that
\begin{eqnarray}
k_p & = & ad\, w_p(h_0) + f_p  \label{e2.9a}  \\
ad\; h_0 (k_p) & = & 0.         \label{e2.9b}
\end{eqnarray}
Equation~(\ref{e2.9a}) and~(\ref{e2.9b}) are solved by the method
of averaging~\cite{LicLie,Moser}. Let $\beta=(\beta_1,\dots ,\beta_n)$
be the coordinates canonically conjugate to the integrals
$b=(b_1,\dots ,b_n)$. Then the flow $\varphi_{h_0}$ generated
by the unperturbed Hamiltonian $h_0$ expressed in the coordinates
$(b,\beta)$ is
\begin{equation}
\varphi_{h_0}(t)
\left( \begin{array}{c} b_0 \\ \beta_0 \end{array} \right) =
\left( \begin{array}{c} b(t) \\ \beta(t) \end{array} \right) =
\left( \begin{array}{c} b_0 \\ \beta_0 +\omega(b_0) t\end{array} \right)
\end{equation}
where
\begin{equation}
\label{e2.10}
\omega(b_0):=\left(\frac{\partial h_0}{\partial b_1}(b_0), \dots,
     \frac{\partial h_0}{\partial b_n}(b_0)  \right)
\end{equation}
gives the frequencies of the unperturbed motion which we assume to be
independent over the rationals (non--resonant) for the given $b_0$,
i.~e.~$c\cdot\omega(b_0)=0\Rightarrow c=0$ for any $c=(c_1,\dots ,c_n)$
with integer $c_j$. Let $g$ be any function on phase space which has the
Fourier decomposition
\begin{equation}
g(b,\beta)=\sum_{c\in {\bf Z}^n} g(b)_c e^{ic\cdot\beta}
\end{equation}
and define the phase space functions
\begin{eqnarray}
\overline{g} & := & \lim_{T\to\infty} \frac{1}{T} \int_0^T
dt\, \varphi_{h_0}(-t)^*g  \label{e2.11}  \\
s(g) & := & \lim_{T\to\infty} \frac{1}{T} \int_0^T dt\,
\int_0^t ds\, \left( \varphi_{h_0}(-s)^*g -\overline{g}\right) \label{e2.12}
\end{eqnarray}
then one finds
\begin{eqnarray}
\overline{g} & = & g(b)_0   \label{e2.13}  \\
s(g) & = & \sum_{c\in{\bf Z}^n-\{0\}} \frac{g(b)_c}{ic\cdot\omega(b)}
e^{ic\cdot\beta}    \label{e2.14}
\end{eqnarray}
and it is easy to see that
\begin{eqnarray}
\overline{g} & = & ad\, (s(g))(h_0) + g  \label{e2.15}  \\
ad\, h_0(\overline{g}) & = & 0.   \label{e2.16}
\end{eqnarray}
With this construction we choose now in~(\ref{e2.9a}) for $w_p$:
\begin{equation}
\label{e2.17}
w_p=s(f_p)
\end{equation}
then $k_p=\overline{f_p}$ commutes with $h_0$ as desired. The important
point to note here is that we have formulated the basic (averaging)
constructions~(\ref{e2.11}) and~(\ref{e2.12}) necessary to
solve~(\ref{e2.9a}) and~(\ref{e2.9b}) in a coordinate free way.
This geometric description using the time average is immediately
suitable for generalization
to self adjoint operators, i.~e.~to quantum mechanics.
It should be noted, however, that (\ref{e2.17}) is not the only
possible solution of~(\ref{e2.9a}) and~(\ref{e2.9b}) since
\begin{equation}
\label{e2.18}
w_p^\prime := w_p + v_p
\end{equation}
is also a solution of our problem as long as
\begin{equation}
\label{e2.19}
ad\; v_p (h_0) = 0.
\end{equation}
This nonuniqueness will also emerge in the quantum mechanical
setting since it is also present in Kato's rigorous exposition
of Rayleigh--Schr\"odinger perturbation theory~\cite{Kato}.
\section{Quantum Poincar\'e--von Zeipel perturbation theory and averaging}
\label{s3}
Now we shall develop the quantum mechanical analogue of the classical
theory presented in the previous section. For this purpose we use
capital latin letters to denote operators on some Hilbert space $\Gamma$.
Let $H_0$ be the unperturbed Hamiltonian operator which is assumed to be
diagonalized in some basis and let
\begin{equation}
\label{e3.1}
H(\epsilon):=\sum_{p=0}^\infty \frac{\epsilon^p}{p!} H_p
\end{equation}
be the perturbed Hamiltonian. Here we do not impose any
conditions (like e.~g.~boundedness) on the perturbations $H_p,\; p\ge 1$,
but proceed purely on a formal level.
A mathematically rigorous justification of each step is
notoriously intricate and will not be attempted here because
it would completely obscure the basic ideas of the method.
Since ultimately we shall prove the equivalence
of this method to the usual Rayleigh--Schr\"odinger perturbation theory
the conditions of the latter theory needed to guarantee
convergence~(see e.~g.~\cite{ReeSim}) will be sufficient to
make the quantum Poincar\'e--von Zeipel
perturbation theory convergent as well.
In analogy to the classical situation we seek a self adjoint
generator (operator)
\begin{equation}
\label{e3.2}
W(\epsilon):=\sum_{p=0}^\infty \frac{\epsilon^p}{p!} W_{p+1}
\end{equation}
such that $-W(\epsilon)$
induces the unitary flow $\Xi(\epsilon)$ with
``time" $\epsilon$, i.e.
\begin{equation}
\label{e3.2a}
\frac{d}{d\epsilon} \Xi (\epsilon) =
\frac{i}{\hbar} W(\epsilon) \Xi(\epsilon),
\qquad \Xi (0) = {\bf 1}.
\end{equation}
Then $\Phi(\epsilon) := \Xi(\epsilon)^{-1}$
is the unique solution of the
initial value problem
\begin{eqnarray}
\frac{d}{d\epsilon}\Phi(\epsilon)^* & = &
AD\; W(\epsilon) \circ \Phi(\epsilon)^*
\label{awpPa}  \\
\Phi (0) & = & {\bf 1}
\label{awpPb}
\end{eqnarray}
where $\Phi(\epsilon)^*$ acts
on any operator $A$ via
\begin{equation}
\label{e3.3}
\Phi(\epsilon)^*A:= \Phi(\epsilon)^{-1}\, A
\, \Phi(\epsilon)
\end{equation}
and where
$AD\, F(G)$
is now defined   as
\begin{equation}
\label{e3.4}
AD\, F(G) := \frac{i}{\hbar} [ F, G]
\end{equation}
for any two operators $F,G$ (again we omit the technical details necessary
to make~(\ref{e3.4}) well defined for unbounded operators)
and as in the classical case
$(AD\, W(\epsilon) )^p:= AD\, W(\epsilon)\circ\cdots\circ AD\, W(\epsilon)$
($p$ times).

As in the classical case it will be useful to expand $\Phi^*$ in terms of
$\epsilon$--independent operators $T_p$:
\begin{equation}
\label{e3.4a}
\Phi(\epsilon)^*=\sum_{p=0}^\infty\frac{\epsilon^p}{p!} T_p
\end{equation}
where the $T_p$ are then recursively defined through $T_0={\bf 1}$ and
\begin{equation}
\label{e3.4b}
T_{p+1}=\sum_{l=0}^p {p \choose l} AD\, W_{l+1}\circ T_{p-l}.
\end{equation}
Note that $\Phi(\epsilon)^*$ and thus the $T_p$ act on
operators whereas $\Phi(\epsilon)$ itself is a transformation
on Hilbert space which can also be expanded as
\begin{equation}
\label{e3.4c}
\Phi(\epsilon) =\sum_{p=0}^\infty \frac{\epsilon^p}{p!}\Phi_p
\end{equation}
and where the following recursive relation for the $\Phi_p$ can
be derived from (\ref{e3.2a}):
\begin{equation}
\label{e3.4d}
\Phi_{p+1} = -\frac{i}{\hbar}\sum_{l=0}^p
{p \choose l} \Phi_{p-l} W_{l+1}
\end{equation}
and $\Phi_0 = {\bf 1}$.

Transforming the perturbed Hamiltonian $H(\epsilon)$ with
$\Phi(\epsilon)$ gives a new Hamiltonian
\begin{equation}
\label{e3.5}
K(\epsilon):=\Phi(\epsilon)^* H(\epsilon)
=\Phi(\epsilon)^{-1} H(\epsilon) \Phi(\epsilon)
\end{equation}
which is assumed to be analytic in $\epsilon$:
\begin{equation}
\label{e3.6}
K(\epsilon)=\sum_{p=0}^\infty\frac{\epsilon^p}{p!} K_p
\end{equation}
and for which one finds
\begin{eqnarray}
K_0 & = & H_0   \label{e3.7} \\
K_p & = & AD\, W_p (H_0) + F_p,\qquad p\ge 1 \label{e3.8}
\end{eqnarray}
with $F_1=H_1$ and for $p\ge 2$
\begin{equation}
\label{e3.9}
F_p:= H_p +\sum_{l=0}^{p-2} {p-1 \choose l} (AD\, W_{l+1} (K_{p-l-1})
+T_{p-l-1} H_{l+1} ).
\end{equation}
All these equations are exactly analogous to the classical case but it
is to be emphasized that they are perfectly well defined operator
equations. But how are we to choose $W$ now? If we could diagonalize
$K(\epsilon)$ up to a given finite order in $\epsilon$, we could
read off its eigenvalues and eigenvectors to that order as well,
but this means that we have found the eigenvalues and the eigenvectors
of the perturbed Hamiltonian $H(\epsilon)$ since by~(\ref{e3.5})
$H$ and $K$ are unitarily equivalent. Before we write this out in
formulae let us first see how we can diagonalize $K$ order by
order using the method of quantum averaging.
It is obvious that the necessary and sufficient condition for
diagonalization is the equivalent of~(\ref{e2.9b}), i.~e.~requiring
\begin{equation}
\label{e3.10}
AD\, H_0 (K_p) = 0, \qquad p\ge 1
\end{equation}
means that all $K_p$ commute with $H_0$ and thus $H_0$ and $K_p$
can be diagonalized simultaneously, hence $K$ can be made diagonal
to any finite order $N$ in $\epsilon$. Consequently, in order to
diagonalize $K$ to any finite order we need to solve the quantum analogue
of~(\ref{e2.9a}) and~(\ref{e2.9b}), i.~e.
\begin{eqnarray}
K_p & = & AD\, W_p(h_0) + F_p  \label{e3.11}  \\
AD\; H_0 (K_p) & = & 0.         \label{e3.12}
\end{eqnarray}
This, too, is done analogous to the classical case. Let
$\Phi_{H_0}$ be the unitary flow generated by the unperturbed
Hamiltonian $H_0$ such that for any operator $G$
\begin{equation}
\label{defgt}
\Phi_{H_0}(-t)^* \, G =
\exp(-\frac{i}{\hbar}tH_0)\, G \, \exp(\frac{i}{\hbar}tH_0) =: G(t)
\end{equation}
where the last equation introduces a simplified notation.
Suppose now that $G$ is such that
\begin{eqnarray}
\overline{G} & := & \lim_{T\to\infty} \frac{1}{T} \int_0^T
dt\, G(t)  \label{e3.13}  \\
S(G) & := & \lim_{T\to\infty} \frac{1}{T} \int_0^T dt\,
\int_0^t ds\, \left( G(s) -\overline{G}\right) \label{e3.14}
\end{eqnarray}
exist and such that
\begin{equation}
\label{e3.15}
\lim_{T\to\infty} \frac{G(T)-G}{T} = 0
\end{equation}
then it follows that
\begin{eqnarray}
\overline{G} & = & AD\, (S(G))(H_0) + G  \label{e3.16}  \\
AD\, H_0(\overline{G}) & = & 0.   \label{e3.17}
\end{eqnarray}
We first prove~(\ref{e3.17}):
\begin{eqnarray*}
AD\, H_0 (\overline{G}) & = &
\frac{i}{\hbar}[H_0, \overline{G}]  =  \lim_{T\to\infty}\frac{1}{T}
\int_0^T \frac{i}{\hbar}[H_0, G(t)] dt  \\
\mbox{} & = & - \lim_{T\to\infty}\frac{1}{T}
\int_0^T \frac{d}{dt}G(t) dt =
 - \lim_{T\to\infty}\frac{G(T)-G}{T} = 0
\end{eqnarray*}
by assumption. Moreover
\begin{eqnarray*}
AD\, (S(G))(H_0) & = &
\lim_{T\to\infty}\frac{1}{T}
\int_0^T dt\int_0^t ds\, \frac{i}{\hbar}[G(s)-\overline{G}, H_0]  =
\lim_{T\to\infty}\frac{1}{T}
\int_0^T dt\int_0^t ds\, \frac{i}{\hbar}[G(s), H_0]  \\
\mbox{} & = &  \lim_{T\to\infty}\frac{1}{T}
\int_0^T dt \int_0^t ds\, \frac{d}{ds}\, G(s) =
 \lim_{T\to\infty}\frac{1}{T}
\int_0^T dt \, \left( G(t) - G\right) \\
\mbox{} & = & \overline{G} - G
\end{eqnarray*}
which proves~(\ref{e3.16}). Equations~(\ref{e3.13})-(\ref{e3.17}) are
the exact quantum analogue of the classical averaging technique
with the noteworthy absence of any non--resonance condition.
Thus for any $p\ge 1$ equation~(\ref{e3.11}) and~(\ref{e3.12}) are
successively solved by
\begin{eqnarray}
W_p & = & S(F_p)  \label{e3.18}  \\
K_p & = & \overline{F_p}.   \label{e3.19}
\end{eqnarray}
Using~(\ref{e3.9}) we may simplify $\overline{F_p}$ by noting that
\begin{equation}
\overline{AD \, W_{l+1} (\overline{F_{p-l-1}})}
= AD\,\overline{W_{l+1}}(\overline{F_{p-l-1}})
\end{equation}
and with~(\ref{e3.18}) we have
$\overline{W_{l+1}} = \overline{S(F_{l+1})}$.
Assuming continuity of the maps $S$ and $\overline{\cdot}$ one can
formally show that $\overline{S(B)}=S(\overline{B})$ for any operator
$B$ for which $S(B)$ and $\overline{B}$ exist. On the other hand it
is evident that $S(\overline{B})=0$. Putting these things together
shows that
\begin{equation}
\label{e3.20}
\overline{AD \, W_{l+1} (\overline{F_{p-l-1}})} = 0
\end{equation}
such that the expression~(\ref{e3.19}) for $K_p$ does not contain
contributons arising from averaging the terms $AD\, W_{l+1} (K_{p-l-1})$
in the expression~(\ref{e3.9}) for the $F_p$.

Let us summarize what we have done so far: Given a perturbed
Hamiltonian $H(\epsilon)=\sum_{p=0}^\infty \frac{\epsilon^p}{p!} H_p$
we have shown that chosing $W_p=S(F_p)$ in
$W(\epsilon)=\sum_{p=0}^\infty \frac{\epsilon^p}{p!} W_{p+1}$
leads to
\begin{equation}
\label{e3.21}
K(\epsilon)=H_0+ \sum_{p=1}^\infty \frac{\epsilon^p}{p!} \overline{F_p} =
\Phi(\epsilon)^{-1}\, H(\epsilon)
\, \Phi(\epsilon)
\end{equation}
and the $\overline{F_p}$ all commute with the unperturbed operator $H_0$
and are given by
\begin{eqnarray}
\overline{F_1} & = & \overline{H_1},  \\
\overline{F_p} & = & \overline{H_p} + \sum_{l=0}^{p-2}
{p-1 \choose l} \overline{T_{p-l-1} H_{l+1}}, \qquad p\ge 2.
\end{eqnarray}
The first few terms in the expansion (\ref{e3.21}) are
\begin{equation}
\begin{array}{cc}
F_0 = 0, & K_0 = H_0, \\
F_1 = H_1, & K_1 = \overline{H_1},  \\
F_2 = H_2 + \frac{i}{\hbar}[W_1, K_1+H_1], &
K_2 = \overline{H_2} + \frac{i}{\hbar}\overline{[W_1, H_1]},
\end{array}
\label{e3.23}
\end{equation}
where $W_1=S(H_1)$.
Hence, $\sum_{p=1}^N \frac{\epsilon^p}{p!} \overline{F_p}$ and $H_0$
can be simultaneously diagonalized for any finite $N$. Let us introduce
the following notation
\begin{eqnarray}
K^N(\epsilon) & = & \sum_{p=0}^N \frac{\epsilon^p}{p!} \overline{F_p}
\label{e3.24}  \\
\Phi^N(\epsilon) & = & \sum_{p=0} \frac{\epsilon^p}{p!} \Phi_p
\label{e3.25a}
\end{eqnarray}
which implies
\begin{equation}
\label{e3.26}
K^N(\epsilon) = (\Phi^N(\epsilon))^{-1} \, H(\epsilon)
\, \Phi^N(\epsilon) + O(\epsilon^{N+1}).
\end{equation}
Let
\begin{equation}
\label{e3.27}
K^N(\epsilon) |j\rangle^N(\epsilon) =
E^N_j(\epsilon) |j\rangle^N(\epsilon)
\end{equation}
and
\begin{equation}
\label{e3.29}
H(\epsilon) |j\rangle (\epsilon) =
E_j(\epsilon) |j\rangle (\epsilon)
\end{equation}
Evidently
\begin{equation}
\label{e3.28}
H(\epsilon) \Phi^N(\epsilon) |j\rangle^N (\epsilon) =
E^N_j(\epsilon) \Phi^N(\epsilon) |j\rangle^N (\epsilon)
+ O(\epsilon^{N+1}),
\end{equation}
i.~e.~the eigenvalues $E_j(\epsilon)$ and
eigenvectors $|j\rangle(\epsilon)$
of the perturbed Hamiltonian $H(\epsilon)$ are approximated as follows
\begin{eqnarray}
E_j(\epsilon) & = & E_j^N(\epsilon) + O(\epsilon^{N+1}) \label{e3.30} \\
|j\rangle(\epsilon) & = &
\Phi^N(\epsilon) |j\rangle^N (\epsilon)
+ O(\epsilon^{N+1}).  \label{e3.31}
\end{eqnarray}
Hence, we have used the quantum analogue of the averaging method
to construct a
quantum mechanical perturbation theory. Just as in the classical
case, however, the solutions for $W_p$ constructed here are not
the only ones. One encounters the same nonuniqueness as
given by (\ref{e2.18}) and (\ref{e2.19}) in the classical
case.
\section{Examples: Discrete spectra}
\label{s4}
\subsection{General second order terms}
In this section we will apply the theory developed in section~\ref{s3}
to the case of a Hamiltonian $H_0$ which is assumed to have a purely
discrete spectrum with finite degeneracy
\begin{equation}
\label{e4.1}
H_0=\sum_{j; \; \alpha\in D_j} |j,\alpha\rangle E^0_j \langle\alpha,j|
\end{equation}
where the sum over $\alpha$ runs over
$D_j:=\{1,\dots, d_j=dim(Eig(H_0, E^0_j))\}$. For any self adjoint $G$ one
then obtains:
\begin{eqnarray}
\overline{G} = \sum_{j;\; \alpha,\beta\in D_j}
|j,\alpha\rangle\langle\alpha,j|G|j,\beta\rangle\langle\beta, j|
\label{e4.2}  \\
S(G)=\frac{\hbar}{i}\sum_{j\neq k;\; \alpha\in D_j,\beta\in D_k}
|j,\alpha\rangle
\frac{\langle\alpha,j|G|k,\beta\rangle}{E^0_j-E^0_k}
\langle\beta, k|.
\label{e4.3}
\end{eqnarray}
Using~(\ref{e3.23}), (\ref{e4.2}), and~(\ref{e4.3}) one finds after
straightforward calculations
\begin{eqnarray}
K^2(\epsilon) & = &
\sum_{j;\; \alpha,\beta\in D_j}
|j,\alpha\rangle
\left( E^0_j + \epsilon \langle\alpha,j|H_1|j,\beta\rangle
\right) \langle\beta,j|
\label{e4.4} \\
\mbox{} &+&
\sum_{j;\; \alpha,\beta\in D_j}
|j,\alpha\rangle
\left(
\epsilon^2\left\{\frac{\langle\alpha,j|H_2|j,\beta\rangle}{2} +
\sum_{j\neq k;\; \gamma\in D_k}
\frac{\langle\alpha,j|H_1|k,\gamma\rangle
\langle\gamma,k|H_1|j,\beta\rangle}{E^0_j-E^0_k}
\right\}\right)
\langle\beta,j|. \nonumber
\end{eqnarray}
Consequently the eigenvalues $E^2_{j,\alpha}(\epsilon)$ of $K^2(\epsilon)$
are determined as solutions of the secular equation of $d_j$--dimensional
matrices:
\begin{equation}
\det\left( E^2 \delta_{\alpha\beta}
- \langle\alpha,j|K^2(\epsilon)|j,\beta\rangle  \right) = 0
\label{e4.5}
\end{equation}
which coincides with the usual Rayleigh--Schr\"odinger result.
The corresponding
eigenvectors of $K^2(\epsilon)$ are then
\begin{equation}
\label{e4.12}
|j,\alpha\rangle^2(\epsilon) =
\sum_{\beta\in D_j} c^{j(2)}_{\alpha\beta}(\epsilon)|j,\beta\rangle
\end{equation}
and one has for the eigenvector $|j,\alpha\rangle(\epsilon)$
of $H(\epsilon)$
\begin{equation}
\label{e4.13}
|j,\alpha\rangle(\epsilon) =
\left({\bf 1}-\epsilon \frac{i}{\hbar} W_1
+\frac{\epsilon^2}{2}\left(
(\frac{i}{\hbar}W_1)^2-
\frac{i}{\hbar}W_2\right)\right)|j,\alpha\rangle^2(\epsilon)
+ O(\epsilon^3)
\end{equation}
with
\begin{equation}
\label{e4.14}
W_1=\frac{\hbar}{i}\sum_{j\neq k;\; \alpha\in D_j, \beta\in D_k}
|j,\alpha\rangle
\frac{\langle\alpha,j|H_1|k,\beta\rangle}{E^0_j-E^0_k}
\langle\beta, k|
\end{equation}
and $W_2=S(H_2+\frac{i}{\hbar}[W_1,\overline{H_1}+H_1])$
which we shall not write
down here but which inserted into~(\ref{e4.13}) yields the
corrections to the eigenvctors to second order
known from Rayleigh--Schr\"odinger perturbation theory
in the case of non--degenerate spectrum.
In this case
we see
from~(\ref{e4.12}) that the eigenvectors of $K^2(\epsilon)$
and $H_0$ coincide: $|j\rangle^2(\epsilon)=|j\rangle$. In fact,
since
\begin{equation}
\label{e4.7}
K^N(\epsilon) = K^{N-1}(\epsilon) + \frac{\epsilon^N}{N!} K_N
\end{equation}
and $[H_0, K_N]=0\;\;\forall N$ we have by induction
\begin{equation}
\label{e4.7a}
|j\rangle^N(\epsilon)=|j\rangle \qquad  \forall N
\end{equation}
in the non--degenerate case.

The general equivalence of the quantum version of the
Poincar\'e--von Zeipel perturbation theory to the standard
Rayleigh--Schr\"odinger
perturbation theory will be proven in section~\ref{s5}.
The fact that the corrections to the eigenvalues
in the Poincar\'e--von Zeipel theory are
derived from an averaging procedure
may, however, provide some computational advantage
since it gives the sums over intermediate states so common
to standard perturbation theory in closed form. For example,
the second order term in the non--degenerate case is given by
\begin{equation}
\label{clofor}
\frac{1}{2} \langle j|H_2|j\rangle + \sum_{j\neq k}
\frac{|\langle j|H_1|k\rangle|^2}{E^0_j-E^0_k}
=\lim_{T\to\infty}\frac{1}{T} \int_0^T dt\,
\left( \frac{1}{2}H_2(t) + \frac{i}{\hbar}[W_1, H_1](t) \right).
\end{equation}
\subsection{Example 1: Anharmonic oscillator}
For the nondegenerate case we shall illustrate the method in the example
of the harmonic oscillator
$H_0=\frac{1}{2}\left(-\frac{d^2}{dx^2}+x^2\right)$ with
cubic perturbation $H_1=\frac{1}{4}x^4,\; H_p=0,\; p\ge 2$
(anharmonic oscillator with $\hbar=1$)
where the quantum Poincar\'e--von Zeipel method will
permit us to compute the sums
over intermediate states and
the corrections up to $O(\epsilon^2)$ without much effort. All
calculations are straightforward if we use the operators
\begin{equation}
\label{defa}
a:=\frac{1}{\sqrt{2}}\left(\frac{d}{dx} + x\right)  \qquad
a^\dagger := \frac{1}{\sqrt{2}} \left(-\frac{d}{dx} + x\right)
\end{equation}
for which one finds
\begin{equation}
\label{at}
a(t)= e^{it} a  \qquad
a^\dagger(t) = e^{-it} a^\dagger
\end{equation}
such that
\begin{eqnarray*}
H_1(t) & = & \frac{1}{16}\left( e^{4it} a^4 +e^{-4it} (a^\dagger)^4 +
              4e^{2it} aH_0a + 4e^{-2it} a^\dagger H_0 a^\dagger \right)
              + \frac{3}{8} (H_0)^2 + \frac{2}{32}  \\
\overline{H_1} & = & \frac{3}{8} (H_0)^2 + \frac{2}{32}  \\
W_1 & = & \frac{1}{64i}\left( (a^\dagger)^4 - a^4 +
          8 a^\dagger H_0 a^\dagger - 8 a H_0 a \right)   \\
W_1(t) & = & \frac{1}{64i}\left(e^{-4it}(a^\dagger)^4 - e^{4it}a^4 +
          8e^{-2it} a^\dagger H_0 a^\dagger - 8e^{2it} a H_0 a \right)   \\
K_2 & = & \frac{1}{2^9}\left( [(a^\dagger)^4, a^4] +
           32 [ a^\dagger H_0 a^\dagger, a H_0 a ] \right)   \\
E^2_j(\epsilon) & = & j +\frac{1}{2} +
             \epsilon\left( \frac{3}{8} (j^2 + j) + \frac{3}{16}\right) -
        \epsilon^2\left( \frac{17}{64} j^3 + \frac{51}{128} j^2 +
                         \frac{59}{128} j + \frac{21}{128} \right).
\end{eqnarray*}
This formula for the correction to the eigenvalues was first derived
by Heisenberg~\cite{Heisen} and is also reproduced by
Kummer using his normal form approach~\cite{Kumm71}.
\subsection{Example 2: Henon--Heiles system}
To illustrate the method for the degenerate case we apply it
to the two--dimensional Henon--Heiles system whose unperturbed
Hamiltonian is (we choose again $\hbar=1$)
\begin{equation}
H_0 =- \frac{1}{2} \left(
\frac{d^2}{dx^2_1}
+\frac{d^2}{dx^2_2}
                \right)
+\frac{(x_1)^2+(x_2)^2}{2}
=: H_{01} + H_{02} =: N_1 +\frac{1}{2} + N_2 +\frac{1}{2}
\end{equation}
where the $N_j,\; j=1,2$ are the number operators and the perturbation is
\begin{equation}
H_1(\alpha,\beta):= \alpha (x_1)^2 x_2 + \beta (x_2)^3,
\qquad H_p=0 \;\; if \;\; p\ge 2.
\end{equation}
It is only for convenience that we have chosen the ``degenerate"
case (i.~e.~equal frequencies for the two one--dimensional
oscillators in $H_0$). The method is completely
oblivious to that distinction.
In this example we treat $\alpha$ and $\beta$ as one perturbation
parameter in the sense that $\alpha=\epsilon\tilde\alpha$,
$\beta=\epsilon\tilde\beta$ and $\epsilon$ is the single
perturbation parameter which we set equal to one at the end of
the calculation. As in the case of the anharmonic oscillator
it is very convenient to use the operators
\begin{equation}
\label{defa12}
a_j: = \frac{1}{\sqrt{2}}\left(\frac{d}{dx_j} + x_j\right)  \qquad
a^\dagger_j := \frac{1}{\sqrt{2}} \left(-\frac{d}{dx_j} + x_j\right)
\end{equation}
which also evolve according to
\begin{equation}
\label{at12}
a_j(t)= e^{it} a_j  \qquad
a^\dagger_j(t) = e^{-it} a^\dagger_j .
\end{equation}
This yields
\begin{eqnarray}
H_1(t) & = & \frac{1}{2\sqrt{2}}
\left\{
e^{3it} \left(
             \alpha (a_1)^2 a_2 + \beta (a_2)^3
              \right) +
e^{-3it} \left(
             \alpha (a_1^\dagger)^2 a_2^\dagger + \beta (a_2^\dagger)^3
             \right)   \nonumber
\right. \\
\mbox{} & + &
e^{it} \left(
             2\alpha H_{01} a_2 + \alpha (a_1)^2 a_2^\dagger +
             2\beta H_{02} a_2 + \beta (a_2)^2 a_2^\dagger
              \right)
\\
\mbox{} & + &
\left.
   e^{-it}  \left(
                   \alpha (a_1^\dagger)^2 +
                  2\alpha H_{01} a_2^\dagger + \beta (a_2^\dagger)^2 a_2
                 +2\beta H_{02} a_2^\dagger
                \right)
\right\}
\nonumber   \\
\overline{H_1} & = & 0
\end{eqnarray}
and
\begin{eqnarray}
W_1 & = & \frac{1}{2\sqrt{2}i}
\left\{
  \frac{1}{3}\left(
          \alpha (a_1^\dagger)^2 a_2^\dagger + \beta (a_2^\dagger)^3
                   \right)
-\frac{1}{3} \left(
     \alpha (a_1)^2 a_2 + \beta (a_2)^3
                     \right)
\right. \nonumber \\
\mbox{} & + & \alpha (a_1^\dagger)^2 +
    2\alpha H_{01} a_2^\dagger + \beta (a_2^\dagger)^2 a_2
    +2\beta H_{02} a_2^\dagger
\\
\mbox{} &-&  \left.  \left(
        2\alpha H_{01} a_2 + \alpha (a_1)^2 a_2^\dagger +
        2\beta H_{02} a_2 + \beta (a_2)^2 a_2^\dagger
          \right)   \right\}
\nonumber
\end{eqnarray}
{}From this one can read off $W_1(t)$ and obtains after some
tedious but straightforward calculations (which we have executed
with the help of the symbolic computation language MAPLE)
\begin{eqnarray}
K_2(\alpha,\beta) & = &
 - \left ({\frac {4\,N_{1}\,N_{2}}{3}}+{\frac {
5}{12}}+{\frac {5\,N_{1}^{2}}{6}}
+(a_1)^2 (a^\dagger_2)^2 + (a_2)^2 (a^\dagger_1)^2
+{\frac {2\,N_2}{3}}+{\frac {3\,N_{1}}{2}}\right )\alpha
^{2}  \nonumber \\ \mbox{}
& - & \left ({\frac {15\,N_{2}}{2}}+\frac{11}{4}+{\frac {15\,N_{2}^{2}}{2}}
\right )\beta^{2} \\ \mbox{}
 & - &\left (\frac{3}{2}+6\,N_{1}\,N_{2}
-{\frac {(a_1)^2 (a^\dagger_2)^2 + (a_2)^2 (a^\dagger_1)^2}{2}}
+3\,N_{2}+3\,N_{1}\right ) \alpha \beta.
\nonumber
\end{eqnarray}
Keeping in mind that we set $\epsilon=1$ and that
$K_1=\overline{H_1}=0$ we now have to find the
eigenvalues $E^2_{(k,\kappa)}(\alpha, \beta)$ of
$K^2(\alpha, \beta)=H_0 +\frac{1}{2}K_2$ which will give us
the correct eigenvalues of $H_0 + H_1$ up to second order.
Let $k=0$ denote the ground state (no degeneracy: $(0,1)$), $k=1$ the
first exited state (double degeneracy: $(1,1),(1,2)$), and $k=3$ the second
exited state (triple degeneracy: $(2,1),(2,\pm)$)
then we find the following eigenvalues
\begin{eqnarray}
E^2_{(0,1)}(\alpha,\beta)  & = &
1-{\frac {11\,\beta^{2}}{8}}-{\frac {5\,\alpha^{2}}{24}}-{\frac {3\,
\beta\,\alpha}{4}}
\nonumber \\
E^2_{(1,1)}(\alpha,\beta) & = &
2-{\frac {11\,\beta^{2}}{8}-\frac{11\,\alpha^{2}}{8}-\frac{9\,
\beta\,\alpha}{4}}
\nonumber \\
E^2_{(1,2)}(\alpha,\beta) & = &
2-{\frac {71\,\beta^{2}}{8}}-{\frac {13\,\alpha^{2}}{24}}-{\frac {9\,
\beta\,\alpha}{4}}
\\
E^2_{(2,1)}(\alpha,\beta) & = &
3-{\frac {71\,\beta^{2}}{8}}-{\frac {19\,\alpha^{2}}{8}}-{\frac {27\,
\beta\,\alpha}{4}}  \nonumber \\
E^2_{(2,\pm)}(\alpha,\beta) & = &
3-{\frac {101\,\beta^{2}}{8}}-{\frac {15\,\beta\,\alpha}{4}}-{\frac {17
\,\alpha^{2}}{8}} \pm {\frac {\sqrt {2025\,\beta^{4}-446\,\beta^{2}
\alpha^{2}-16\,\alpha^{3}\beta+41\,\alpha^{4}}}{4}}. \nonumber
\end{eqnarray}
The results for $E^2$ agree with those obtained by
Kummer~\cite{Kumm93} and Ali~\cite{Ali}
(except for the factor of $\alpha^2$ in $E^2_{(2,1)}$ in~\cite{Ali}
which is probably due to a typographical error).
\section{Equivalence to Rayleigh--Schr\"odinger perturbation theory}
\label{s5}
\subsection{Non--degenerate case}
In section~\ref{s4} we have already seen that at least up to second
order the quantum analogue of the Poincar\'e--von Zeipel method
and the standard Rayleigh--Schr\"odinger perturbation theory coincide.
In this section we show that this is indeed true for the
full perturbation expansions. To do this we recall briefly how
the standard Rayleigh--Schr\"odinger expansion is constructed in the
nondegenerate case. With the help of a suitably chosen contour
integral in the complex $E$--plane one can show that the
projector
\begin{equation}
\label{projint}
P_j(\epsilon) =\frac{1}{2\pi i}\oint_{|E-E^0_j|=r} \frac{dE}{E-H(\epsilon)}
\end{equation}
on the $j$th eigenspace of $H(\epsilon)$ is analytic in $\epsilon$
and that for $\epsilon$ sufficiently small
$\langle j| P_j(\epsilon) |j\rangle > 0$, \cite{Kato,ReeSim}.
This projector then gives a normalized eigenvector $|j\rangle(\epsilon)$
of $H(\epsilon)$ to the eigenvalue $E_j(\epsilon)$ via
\begin{equation}
\label{e5.1}
|j\rangle(\epsilon) =
\frac{P_j(\epsilon)|j\rangle}{\sqrt{\langle j|P_j(\epsilon)|j\rangle}}.
\end{equation}
{}From this one obtains
\begin{equation}
\label{e5.2}
E_j(\epsilon)=(\epsilon)\langle j|H(\epsilon)|j\rangle(\epsilon) =
\frac{\langle j|P_j(\epsilon) H(\epsilon) P_j(\epsilon) |j\rangle}{
\langle j|P_j(\epsilon)|j\rangle} .
\end{equation}
Using the expansion for $P_j(\epsilon)$ the right hand side
of~(\ref{e5.2}) then yields an expansion for $E_j(\epsilon)$ in
$\epsilon$ which is the usual Rayleigh--Schr\"odinger perturbation series.

On the other hand it follows from~(\ref{e4.7a}) that
\begin{equation}
\label{e5.3}
K(\epsilon)|j\rangle=E_j(\epsilon)|j\rangle
\end{equation}
where $K(\epsilon)=\Phi(\epsilon)^{-1}\, H(\epsilon)\,
\Phi(\epsilon)$ which implies
\begin{equation}
\label{e5.4}
|j\rangle(\epsilon) = \Phi(\epsilon) |j\rangle
\end{equation}
such that
\begin{equation}
\label{e5.5}
P_j(\epsilon) = |j\rangle (\epsilon)(\epsilon)\langle j| =
\Phi(\epsilon)) |j\rangle\langle j| \Phi(\epsilon)^{-1}.
\end{equation}
Inserting~(\ref{e5.5}) in~(\ref{e5.2}) yields
\begin{eqnarray}
E_j(\epsilon) & = &
\frac{\langle j|P_j(\epsilon) H(\epsilon) P_j(\epsilon) |j\rangle}{
\langle j|P_j(\epsilon)|j\rangle}
= \langle j| \Phi(\epsilon)^{-1} \,H(\epsilon)
  \, \Phi(\epsilon)j\rangle   \nonumber \\
\mbox{} & = &
\langle j| K(\epsilon) |j\rangle =
\sum_{p=0}^\infty  \frac{\epsilon^p}{p!} \langle j|K_p |j\rangle
\label{eqpro}
\end{eqnarray}
which proves the equivalence in all orders for the non--degenerate case.
\subsection{Degenerate case}
Suppose now that $H_0|j,\alpha\rangle = E^0_j|j,\alpha\rangle$ with
possible degeneracies $\alpha\in D_j=\{1,\dots ,d_j\}$, that $E^0_j$ is
an isolated point of the spectrum $\sigma(H_0)$, that
$P_j=\sum_{\alpha\in D_j} |j,\alpha\rangle\langle\alpha,j|$ is the
projector on the $j$th eigenspace of $H_0$ and that
$P_j(\epsilon)$ defined as in~(\ref{projint}) exists and is analytic
in $\epsilon$. Then it has been shown that~\cite{Kato,ReeSim}
\begin{equation}
\label{e5.7}
\sigma\left( H(\epsilon)\mid_{Ran\, P_j(\epsilon)} \right) =
\sigma( H(\epsilon)) \cap \{ E \mid |E-E^0_j|<r\}
\end{equation}
(where $A\mid_{Ran\, B}$ means restriction of the operator $A$
to the range of $B$) and that there exists a unitary operator
$U(\epsilon)$ such that
\begin{equation}
\label{e5.8}
P_j(\epsilon) = U(\epsilon) P_j U(\epsilon)^{-1}
\end{equation}
and
\begin{equation}
\label{e5.9}
\tilde{H}(\epsilon) := U(\epsilon)^{-1} H(\epsilon) U(\epsilon)
\end{equation}
satisfies
\begin{equation}
\label{e5.10}
\tilde H (\epsilon) P_j = P_j \tilde H (\epsilon).
\end{equation}
Then $P_j( \tilde H(\epsilon) - E)P_j$ is a finite dimensional operator
analytic in $\epsilon$ and the eigenvalues $E_{j,\alpha}(\epsilon)$
of $H(\epsilon)$ are found as the $d_j$ roots of the equation
\begin{equation}
\label{e5.11}
\det \left( P_j ( \tilde H(\epsilon) - E ) P_j \right) =0.
\end{equation}
The operators $L_j(\epsilon):=P_j U(\epsilon)^{-1}$
and $R_j(\epsilon):=U(\epsilon) P_j$
sandwiching $H(\epsilon)$ in (\ref{e5.11}) satisfy certain
differential equations involving $P_j(\epsilon)$ whose
expansion in terms of $\epsilon$ is known from
(\ref{projint}). With the help of these differential equations
$L_j(\epsilon)$ and $R_j(\epsilon)$ can be expanded in
$\epsilon$ as well.
The $N$th order approximation in the Rayleigh--Schr\"odinger series
for the degenerate case is then obtained by solving~(\ref{e5.11})
where terms of order higher than $N$ are neglected.
We refer the reader to \cite{Kato} for more details on Kato's rigorous
exposition of the usual quantum mechanical
Rayleigh--Schr\"odinger perturbation theory.

As pointed out in \cite{Kato} the unitary transformation $U(\epsilon)$
exists but is not necessarily unique. As we shall see below
this nonuniqueness is equivalent to the nonuniqueness of the
choice of $W_p$ mentioned at the end of section \ref{s3}.

In order to prove
the equivalence with the quantum Poincar\'e--von Zeipel method
we first note that
\begin{equation}
\label{e5.12}
\Phi(\epsilon)^{-1} \, U(\epsilon) P_j =
P_j \Phi(\epsilon)^{-1} \, U(\epsilon).
\end{equation}
In fact, since $[K(\epsilon), H_0]=0$ we may write the orthonormalized
eigenvectors of $K(\epsilon)$ as
$\sum_{\beta\in D_j} u^j(\epsilon)_{\alpha\beta}|j,\beta\rangle$
where $u^j(\epsilon)$ is a $d_j$--dimensional unitary matrix. Hence,
the projector on the space
$\bigoplus_{\alpha\in D_j} Eig(K(\epsilon), E_{j,\alpha}(\epsilon))$
is
\begin{equation}
\label{e5.13}
\sum_{\alpha,\beta,\gamma\in D_j} u^j(\epsilon)_{\alpha\beta}
|j,\beta\rangle\langle\gamma,j|
\left(u^j(\epsilon)\right)^\dagger_{\gamma\alpha} =
\sum_{\beta\in D_j} |j,\beta\rangle\langle\beta,j| = P_j.
\end{equation}
Since
$K(\epsilon)=\Phi(\epsilon)^{-1} H(\epsilon) \Phi(\epsilon)$, one
has
\begin{equation}
\label{e5.14}
P_j(\epsilon) = \Phi(\epsilon)
\, P_j\, \Phi(\epsilon)^{-1}
\end{equation}
which together with (\ref{e5.8}) proves~(\ref{e5.12}).
Equation (\ref{e5.12}) states that
\begin{equation}
Z(\epsilon) := \Phi(\epsilon)^{-1} \, U(\epsilon)
\end{equation}
is a unitary transformation commuting with all $P_j$.
On the other hand, from
\begin{equation}
\Phi(\epsilon) \, K(\epsilon) \, \Phi(\epsilon)^{-1} = H(\epsilon) =
U(\epsilon) \, \tilde H (\epsilon) \, U(\epsilon)^{-1}
\end{equation}
it follows that
\begin{equation}
\label{hzk}
\tilde H (\epsilon) =
Z(\epsilon)^{-1} \, K(\epsilon) \, Z(\epsilon)
\end{equation}
which implies that the roots of equation (\ref{e5.11}) are
identical to the roots of
\begin{equation}
\det \left( P_j (K(\epsilon) - E ) P_j \right) = 0
\end{equation}
and this proves the equivalence
of the two methods for the eigenvalue--expansions
in the presence of degenerate eigenvalues.

Moreover, as can be seen from (\ref{hzk}) the eigenvectors of
$\tilde H (\epsilon)$ are related to those of $K(\epsilon)$
by the unitary transformation $Z(\epsilon)$ which preserves
each eigenspace of $H_0$. It is likely that the nonuniqueness
in the choice of $W_p$ may be exploited to make $Z(\epsilon)$
trivial \cite{MKpriv} but this is still under investigation.
\section{Discussion and conclusion}
\label{s6}
Kummer~\cite{Kumm71} was the first to discuss the averaging method
for quantum systems.
Based on ideas from classical averaging
he constructed a perturbation method called the
normal form approach~\cite{Kumm93} which is equivalent to
``time averaging"~\cite{Kumm71} but instead of using averaging
to solve~(\ref{e3.11}) and~(\ref{e3.12}) it employs algebraic
constructions.

Motivated by the Birkhoff--Gustavson normal form in classical
mechanics Ali~\cite{Ali}  has developed a
quantum analogue of this and his construction yields the same
expansion as that of Kummer.
Working explicitly with an algebra of destruction and creation
operators Eckhardt~\cite{Eckhar} has also constructed a quantum
analogue of the Birkhoff--Gustavson normal form.
A quantization of the classical Birkhoff--Gustavson normal form
has been attempted by Robnik~\cite{Robnik} but this is necessarily
plagued by ordering problems which do not affect our work and
the other contributions cited above (this is only partly true
for~\cite{Eckhar}).

The constructions of Kummer, Ali and Eckhardt have in common
that the existence of the generators of the unitary
transformation has to be assumed or assured by certain additional
conditions. In the present paper the necessary generators $W_p$
are (at least formally)
explicitly constructed.

In fact, it can be shown \cite{MKpriv} that the method
of quantum averaging as presented here provides explicit
solutions in terms of the time averaging integrals
for the algebraic constructions of Kummer.
The algebraic constructions have the advantage
of rigorous validity but lack constructive procedures
needed to execute the algorithm. The approach presented
here has -- apart from its conceptual proximity to the
classical situation -- the advantage of providing
explicit constructions. Due to the analytic character
of these constructions, however, technical problems
which are absent in the algebraic approach may arise.

Using a slightly modified quantum averaging in our sense for the
particular example
of the quantum anharmonic oscillator Ben Lemlih and Ellison~\cite{LemEll}
have derived rigorous error bounds on approximations to the quantum time
evolution.

Their work also contains a suggestion to compare the approximation
to the eigenvalues of this specific problem to the usual perturbative
corrections, i.~e.~Rayleigh--Schr\"odinger theory.
Ali~\cite{Ali} and and Kummer~\cite{Kumm93} have found that in all the
examples they have treated the normal form perturbative results
agree with Rayleigh--Schr\"odinger theory (incidentally this is not
true for Robnik's
expansions~\cite{Robnik}) and
Eckhardt also suggests that the Birkhoff--Gustavson perturbation expansion
is identical to the usual Rayleigh--Schr\"odinger pertubation theory.
Our work then provides an explicit proof of this assertion since,
as Kummer
has shown, his normal form approach is equivalent to the averaging
method in quantum mechanics and we have shown that averaging is
completely equivalent to the Rayleigh--Schr\"odinger theory
yielding the sums over intermediate states in closed forms.

A very important aspect related to this work
(but not discussed here) is the fact,
that just as in classical mechanics a superconvergent perturbation
theory can be constructed with the help of averaging, this can be
done in quantum mechanics as well and yields a perturbation theory
explicitly distinct from the usual Rayleigh--Schr\"odinger
theory~\cite{SUCO,QA2}.

Work is in progress to establish the technical conditions necessary
to put quantum averaging on a rigorous mathematical footing
and to determine how the nonuniqueness of the $W_p$
may be used to trivialize $Z(\epsilon)$.
It may also be possible that standard time--dependent perturbation
techniques (e.~g. sudden aproximation) can be formulated as
analogues of classical time--dependent averaging.
\subsubsection*{Acknowledgements}
I am greatly indebted to M.~Kummer for a critical reading of the
first version of this paper and very instructive remarks
in which he pointed out some errors in the original version.
Helpful discussions with H.~D.~Doebner and D. Mayer are
also gratefully acknowledged.

\end{document}